\documentclass[pdftex,twocolumn,epjc3]{svjour3}          

\RequirePackage[T1]{fontenc}

\smartqed  

\RequirePackage{graphicx}
\RequirePackage{mathptmx}      
\RequirePackage{flushend}
\RequirePackage[numbers,sort&compress]{natbib}
\RequirePackage[colorlinks,citecolor=blue,urlcolor=blue,linkcolor=blue]{hyperref}
\RequirePackage{multirow}
\RequirePackage{morefloats}
\RequirePackage{amssymb,amsmath}
\RequirePackage{cases}
\journalname{Eur. Phys. J. C}

\begin{document}

\title{A new mass limit ($3.61~M_{\odot}$) of strange star admitting CFL equation of state}


\author{K. B. Goswami\thanksref{e1,addr1}
        \and
        A. Saha\thanksref{e2,addr1,addr2}
        \and
        P. K. Chattopadhyay\thanksref{e3,addr1}
        \and
        S. Karmakar\thanksref{e4,addr3}
        }
\thankstext{e1}{e-mail: koushik.kbg@gmail.com}
\thankstext{e2}{e-mail: anirban.astro9@gmail.com}
\thankstext{e3}{e-mail: pkc$_{-}$76@rediffmail.com}
\thankstext{e4}{e-mail: skarma78@rediffmail.com}

\institute{Department of Physics, Coochbehar Panchanan Barma
University, Vivekananda Street, District: Coochbehar, Pin: 736101, West Bengal, India\label{addr1}         
\and
Department of Physics, Alipurduar College, Alipurduar, Pin:736122, West Bengal, India\label{addr2}
\and
Department of Physics, University of North Bengal, Raja Rammohunpur, P.O.-N.B.U., District-Darjeeling, PIN-734013, West Bengal, India \label{addr3}}
\date{Received: date / Accepted: date}

\maketitle

\begin{abstract}
A class of strange star is analyzed in the present article in hydrostatic equilibrium whose state is defined by a CFL phase equation of state. We compare our result with those obtained from MIT bag equation of state for strange quark matter which are regarded as free particles. We note that if we consider quarks to form cooper pair and their description is made by CFL equation of state, the maximum mass of strange star assumes value as high as 3.61 $M_{\odot}$ which is well above the value 2.03 $M_{\odot}$ obtained by considering MIT bag equation of state for massless free quarks. Both the maximum masses are determined by solving TOV equation for different values of strange quark mass $m_s$. Thus inclusion of possibility of quark pair formation in the theory permits us to accommodate a wider class of compact objects like 4U 1820-30, PSR J1614-2230, PSR J0030+0451, PSR J1903+0327, PSR J0740+6620, PSR J0952-0607 and mass of the companion star in GW170817 and GW190814 events in our model. The consideration of such high value of mass is hardly obtainable theoretically from normal strange star models in General Relativity even with fast rotation effect. The object PSR J0952-0607 is found to be the fastest and heaviest pulsar in the disk of Milky Way Galaxy having mass 2.59 $M_{\odot}$ may be predicted in our model as observational evidence supports the existence of strange quark matter in its composition.

keywords: strange star; anisotropy; MIT EoS; CFL EoS; strange quark mass     
\end{abstract}

\section{Introduction}\label{intro}
The objective that strange quark matter ($henceforth~SQM$) may be the true ground state of quantum chromodynamics ($henceforth~QCD$) was first suggested by Witten \cite{Witten} in 1984. The basic concept is that in case of SQM, the energy associated to each baryon might be lower than that of the energy per baryon of most stable nucleus, $^{56}Fe$, making SQM more stable. According to the asymptotic behaviour of QCD, nucleons are split up into the form of quarks and these weakly interacting de-confined quarks constitute a gas. Theories used for the description of hadron to quark phase transition is phenomenological. One such model is bag model \cite{Chodos} which contributes the suitable description of the bulk quark matter and also its confinement in a region of space called a "bag", containing hadronic fields. The bag has a constant, positive potential energy per unit volume. This constant is known as vacuum energy $B$ also termed as bag constant. In the context of MIT bag model, Farhi and Jaffe \cite{Farhi} studied extensively the properties of SQM considering different choices of $B$ and the mass of strange quark ($m_{s}$). They have established a stability window for stable SQM in the ($m_{s}-B$) plane. In the context of General Relativity ($henceforth$ $GR$), a number of authors have used the MIT bag Equation of State ($henceforth$ $EoS$) to study compact stars composed of quark matter qualitatively \cite{Brilenkov,Paulucci,Arbanil,Lugones1,Chowdhury}. If the hypothesis of strange matter is correct, a possibility of new class of compact object made entirely of SQM, called the "Strange Star" should exist. Possible existence of strange stars was studied by many researchers \cite{Haensel,Alcock} and it is confirmed theoretically that stable strange star configurations are possible. These stars are much smaller in size than that of normal neutron stars because they are self bound due to strong interaction and fundamentally different in comparison to the gravitationally bound neutron stars \cite{Haensel,Alcock}.\\ 
There has been a vast theoretical progress in the understanding of SQM. It is well established that in presence of weak attractive interaction arbitrary in nature, the degenerate Fermi systems are unstable. Such problem of instability is solved by considering the formation of cooper pairs in Bose condensate form leading to superconductivity for charged fermions. Similarly at low temperature and sufficiently high baryon number density, the presence of attractive interaction between quarks will lead to quark Cooper pairs and lead to color-superconductivity. It is now well admitted that color flavor locked (CFL) strange matter is the true ground state of strong interaction \cite{Lugones2}. At temperatures much lower than quark chemical potentials ($T<<\mu$), several spontaneous symmetry breaking phases occur. In nature this only happens in the interior of compact objects where matter density exceeds nuclear matter density. The time span of such objects are high enough to equilibrate through weak interactions and to reduce the temperature of it below the chemical potential of quarks. It is expected that at low temperatures and ultra-high densities hadrons may breaks into a degenerate soup of quarks forming Cooper pairs near the Fermi surface which exhibits color superconductivity. The quark Cooper pair formation was noted long before the development of any consistent theory of strong interaction \cite{Ivanenko1,Ivanenko2}. The study of quark Cooper pair formation originated after the pioneering works of Barrois \cite{Barrois1,Barrois2} and Frautschi \cite{Frautschi}. Most of the studies of SQM in both normal and CFL phase initiated the concept of vacuum energy based on the idea of phenomenological bag constant, $B$. But due to the high densities in the vicinity of core of the compact stars/ strange stars it is more practical to demand that the bag constant should be density dependent \cite{Reinhardt}. Several works have been done on density dependent bag constant \cite{Chakrabarty1,Chakrabarty2,Peng,Prasad,Zhu}. Thermodynamics and the EoS of the SQM are different if one consider the density dependent $B$ from those with constant $B$. Our objective here is to study the strange star with density dependent bag constant model and compare it with an exactly solvable Vaidya-Tikekar model \cite{Vaidya,Mukherjee}. In Vaidya-Tikekar model, the $g_{rr}$ metric component extensively depends on the spheroidal parameter $\lambda$ and curvature parameter $R$. Which implies that the physical 3-space associated with the star possess the geometry of a 3-spheroid immersed in a four-dimensional Euclidean flat space. Using V-T ansatz, Sharma {\it et al.} \cite{Sharma1} have shown by fine tuning of $\lambda$ or bag constant $B$, the observed mass radius of a wider range of compact objects may be predicted from theoretical model. Goswami {\it et al.} \cite{Goswami} have obtained recently the radius and maximum mass of strange stars whose interior may be described by MIT bag EoS in V-T model. They also showed that a correlation between the spheroidal parameter $\lambda$ and bag constant $B$ exists so that observed mass and radius of some compact objects may be predicted from their model. The prime goal of the present article is to make a comparative study between quark matter in CFL phase EoS and MIT bag EoS and their astrophysical implications. The thermodynamically obtained pressure density relation has been related with that obtained by solving Einstein Field equations for metric ansatz of Vaidya \& Tikekar.\\ The paper is arranged in the following sequences: in Sect.~\ref{sqm} we have given a brief outline of the thermodynamic properties of SQM and the EoS in MIT bag model. Sect.~\ref{density} consists of density dependent bag model as proposed by previous authors while the origin of density dependent bag constant ($B$) has been discussed on thermodynamic stand point in Sect.~\ref{thermo}. The EoS of quarks in CFL phase is outlined in Sect.~\ref{eos}. In Sect.~\ref{vaidya} the solutions of Einstein Field equations is presented for Vaidya-Tikekar ansatz. Hence the mathematical expressions of energy density and pressure have been obtained. Next we have used the thermodynamically obtained EoS for both CFL and MIT bag EoS and incorporated them in the mathematical model which have stated in Sect.~\ref{physical}. A comparison of both the EoS have been made which also have been discussed in this section. In Sect.~\ref{m-r} we have shown the mass-radius relation of strange stars in CFL and MIT bag EoS. The energy conditions are plotted in Sect.~\ref{ec}. The stability of the model has been studied in view of TOV equation, Herrera cracking concept and the nature of adiabatic index which have been represented in Sect.~\ref{eq}-\ref{ad}. Finally we make concluding remarks on our model by mentioning some of the striking features in Sect.~\ref{conc}.

\section{General properties of SQM}\label{sqm}
SQM is modified as a Fermi gas comprised of massless $u$ and $d$ quarks, massive strange ($s$) quark of mass $m_{s}$ and electrons. The chemical equilibrium between the particles is sustained through the weak interactions
\begin{equation}
d,s \leftrightarrow u+ e + \overline{\nu}_{e}~~~~;~~~ s+u \leftrightarrow u+d. \label{Eq1}
\end{equation} 
The properties of strange matter are determined by their thermodynamic potentials $\Omega_{i}(i=u,d,s,e^{-})$ which are functions of chemical potentials $\mu_{i}$ as well as $m_{s}$ and the strong interaction coupling constant $\alpha_{c}$ \cite{Farhi}. The weak interactions given by Eq.~(\ref{Eq1}) imply that the chemical potentials $\mu_{i}$ should satisfy
\begin{equation}
\mu_{d}=\mu_{s}\equiv \mu~~~;~~~\mu_{u}+\mu_{e}=\mu \label{Eq2}
\end{equation}
and the charge neutrality condition requires
\begin{equation}
\frac{2}{3}n_{u}-\frac{1}{3}n_{d}-\frac{1}{3}n_{s}-n_{e}=0, \label{Eq3}
\end{equation}
where $n_{i}$ is known as the number density of $i$th type particle given by $n_{i}=-(\frac{\partial \Omega_{i}}{\partial \mu_{i}})$. Eqs.~(\ref{Eq2}) and (\ref{Eq3}) imply that there is only one independent chemical potential denoted by $\mu$. The pressure $(p)$, energy density $(\rho)$ and the baryon number density $(n)$ of the quark phase are evaluated from the following relations
\begin{equation}
p=-\sum_{i} \Omega_{i}-B, \label{Eq4}
\end{equation}
\begin{equation}
\rho=\sum_{i} {(\Omega_{i}+\mu_{i}n_{i})}+B, \label{Eq5}
\end{equation}
\begin{equation}
n=\frac{1}{3}(n_{u}+n_{d}+n_{s}). \label{Eq6}
\end{equation}
To obtain the EOS, one has to eliminate the $\mu$ from Eq.~(\ref{Eq4}) and express $\rho$ and $n$ as a function of pressure $p$ using Eqs.~(\ref{Eq5}) and (\ref{Eq6}). Witten \cite{Witten} considered $m_{s}\rightarrow 0$, $\alpha_{c}\rightarrow 0$ in case of neutron star and give the EOS of the quark matter approximately as
\begin{equation}
p=\frac{1}{3}(\rho-4B). \label{Eq7}
\end{equation}

\section{Density dependent Bag Constant}\label{density}
Originally, the bag constant was kept fixed as its free space value in the bag model. Since the de-confinement phase transition depends on both temperature and the number density of baryon of the system, the bag constant may be temperature dependent \cite{Muller} as well as density dependent \cite{Liu,Burgio,Aguirre}. Bag constant which depends on the temperature describes the scenario of heavy-ion collision in the high energy regime of terrestrial laboratories whereas the density dependent bag constant describes the cold compact stars. Prasad and Bhalerao \cite{Prasad} considered three forms of the density dependent $B$ published by three different groups:
\begin{enumerate}
\item They have fitted the results obtained by Liu {\it et al.} \cite{Liu} to express $B$ analytically as
\begin{equation}
B(n)=B(0)exp[-(a_{1}x^{2}+a_{2}x)], \label{Eq8}
\end{equation}
where $n$ is the baryon number density, $x=\frac{n}{n_{0}}$ is normalised number density of baryons. Where $n_{0}=0.17~fm^{-3}$ represents the baryon number density of the ordinary nuclear matter, $a_{1}=0.0125657$, $a_{2}=0.29522$ and $B(0)=114~MeV/fm^{3}$ $={(172~MeV)}^{4}$.
\item Burgio {\it et al.} \cite{Burgio} have presented $B$ in terms of a parametric form given below:
\begin{equation}
B(n)=B_{as}+(B_{0}-B_{as})exp[-\beta x^{2}], \label{Eq9}
\end{equation}
where $B_{as}=38~MeV/fm^{3}$, $B_{0}=200~MeV/fm^{3}$, $\beta=0.14$.
\item Aguirre \cite{Aguirre} has calculated a density dependent $B$ to study SQM in the CFL phase, which is given by
\begin{equation}
B(n)=a+\sum_{i=1}^{5}b_{i}x^{i},~x \leq 9;~~~\\ 
B(n)=\beta exp[-\alpha(x-9)],~x>9, \label{Eq10}
\end{equation}
\end{enumerate}
where $a=291.5906$, $b_{1}=-142.25581$, $b_{2}=39.29997$, $b_{3}=-6.04592$, $b_{4}=0.46817$, $b_{5}=-0.01421$, $\alpha=0.253470705$ and $\beta=19.68764$.

\begin{figure}[ht!]
\centering
\includegraphics[width=8cm]{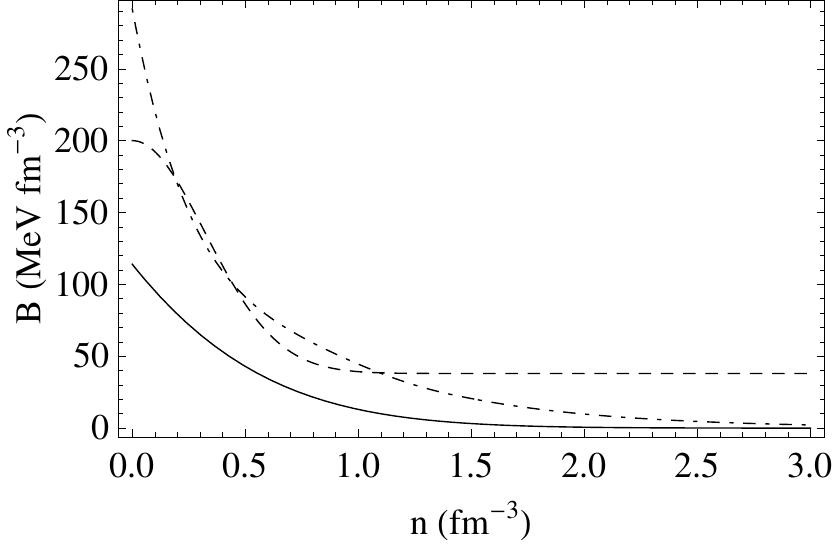}
\caption{Bag constant $B$ vs baryon number density $n$ for the three models. The solid line is for Liu {\it et al.} \cite{Liu}, the dotdashed line is for Aguirre \cite{Aguirre} and the dashed line is for Burgio {\it et al.} \cite{Burgio}}
\label{Fig1}
\end{figure}
Density dependent $B$ given above are displayed in Fig.~\ref{Fig1}. The above three models give the value of $B$ over a broad range of densities exist in the strange star.

\section{Thermodynamics with density dependent B}\label{thermo}
Here we follow the similar mechanism adopted by Zhu {\it et al.} \cite{Zhu}. First we derive from the general ensemble theory the expression for energy and pressure of a system of particles when the particle masses are dependent on density. The partition function is given by
\begin{equation}
\Xi=\sum_{{N_{i}},\alpha} e^{-\beta(E_{N_{i},\alpha}-\sum_{i}\mu_{i}N_{i})}, \label{Eq11}
\end{equation}
where $\beta$ is the reverse temperature, $N_{i}$ and $\mu_{i}$ represent respectively the particle numbers and chemical potentials of $i$th type particle. The microscopic energy $E_{N_{i},\alpha}=f(V,m_i,N_i,\alpha)$ where $V$, $m_i$, $N_i$ and $\alpha$ represent the volume of the system, particle masses, particle numbers and other quantum numbers. The pressure of such system is defined as
\begin{eqnarray}
p&=&\frac{1}{\Xi}\sum_{{N_{i}},\alpha}\left(-\frac{\partial E_{N_{{i},\alpha}}}{\partial V}\right)e^{-\beta(E_{{N_{i},\alpha}}-\sum_{i}\mu_{i}{N_{i}})}\nonumber \\
&& \hspace{-0.2cm} =\frac{1}{\beta}\frac{\partial ln \Xi}{\partial V}=-\frac{\partial (V\Omega)}{\partial V}, \label{Eq12}
\end{eqnarray}
where $\Omega=-\frac{1}{V \beta}ln \Xi$ is the thermodynamic potential density. Which can be written in the functional form as $\Omega=f(T,\mu_i,m_i)$. If the masses of particles are independent of baryon number density $n_{b}=\left(\frac{N}{3V}\right)$ ($N= \sum N_i$), one gets $p=-\Omega$.\\
However, in the situation where particles mass depends on the density or volume, one should have
\begin{equation}
p=-\Omega+n_{b}\frac{\partial \Omega}{\partial n_{b}}. \label{Eq14}
\end{equation}
Now the statistical average for the energy is given by
\begin{eqnarray}
\bar{E}&=&\frac{1}{\Xi}\sum_{{N_{i}},\alpha}E_{N_{i},\alpha}e^{-\beta(E_{{N_{i},\alpha}}-\sum_{i}\mu_{i}{N_{i}})}\nonumber \\
&& \hspace{-0.2cm} =-\frac {\partial}{\partial \beta} ln \Xi +\sum_{i}\mu_{i}{\bar{N_{i}}}, \label{Eq15}
\end{eqnarray}
where
\begin{eqnarray}
\bar{N_{i}}&=&\frac{1}{\Xi}\sum_{{N_{i}},\alpha}N_{i}e^{-\beta(E_{{N_{i},\alpha}}-\sum_{i}\mu_{i}{N_{i}})}\nonumber \\
&& \hspace{-0.2cm} =\frac{1}{\beta}{\left(\frac {\partial}{\partial \mu_{i}} ln \Xi\right)}_{V,T,m_{k}}=-V{\left(\frac{\partial \Omega}{\partial \mu_{i}}\right)}_{T,m_{k}}, \label{Eq16}
\end{eqnarray}
represents average of number of $i$th type particles. Hence, the energy density of such system can be written as:
\begin{eqnarray}
\rho &=&\frac{\bar E}{V}=\frac{\partial (\beta \Omega)}{\partial \beta}+\sum_{i}\mu_{i}n_{i} \nonumber \\
&& \hspace{-0.2cm} =\Omega + \beta \frac{\partial \Omega}{\partial \beta}+\sum_{i}\mu_{i}n_{i} \nonumber \\
&& \hspace{-0.2cm} =\Omega + \sum_{i}\mu_{i}n_{i}-T \frac{\partial \Omega}{\partial T}. \label{Eq17}
\end{eqnarray}
where $n_{i}$ is the number density of particle type $i$ and it is given by
\begin{equation}
n_{i}= \frac{\bar N_{i}}{V}=-{\left(\frac{\partial \Omega}{\partial \mu_{i}}\right)}_{T,m_{k}} \label{Eq18}
\end{equation}
The energy of the system in the MIT bag model in microscopic regime is expressed as
\begin{equation}
E_{N_{i}, \alpha}^{Bag}=E_{N_{i}, \alpha}+BV. \label{Eq19}
\end{equation}
The partition function becomes
\begin{equation}
\Xi^{Bag}=\sum_{N_{i},\alpha}e^{-\beta(E_{N_{i},\alpha}+BV-\sum_{i}\mu_{i}N_{i})}=\Xi e^{-\beta BV}. \label{Eq20}
\end{equation}
Therefore
\begin{equation}
n_{i}^{Bag}=\frac{\bar N_{i}}{V}=-{\left(\frac{\partial (\Omega+B)}{\partial \mu_{i}}\right)}_{T,m_{k},E_{N_{i},\alpha},B}, \label{Eq21}
\end{equation}
\begin{eqnarray}
p^{Bag}&=&\frac{1}{\beta}\frac{\partial ln\Xi^{Bag}}{\partial V}=\frac{1}{\beta}\frac{\partial (ln\Xi-\beta BV)}{\partial V} \nonumber \\
&& \hspace{-0.2cm} =-(\Omega+B)-V\frac{\partial(\Omega+B)}{\partial V}, \label{Eq22}
\end{eqnarray}
\begin{equation}
\rho^{Bag}=\Omega +B+\sum_{i} \mu_{i}n_{i}-T\frac{\partial (\Omega+B)}{\partial T}. \label{Eq23}
\end{equation}
Now if we consider that mass of particle does not depend on the number density of baryon and the parameter $B$ depends only on energy density, then Eqs.~(\ref{Eq21})-(\ref{Eq23}) reduces to 
\begin{equation}
n_{i}^{Bag}=-\left(\frac{\partial \Omega}{\partial \mu_{i}}\right), \label{Eq24}
\end{equation}
\begin{equation}
p^{Bag}=-(\Omega+B)+n_{b}\frac{\partial B}{\partial n_{b}}, \label{Eq25}
\end{equation}
\begin{equation}
\rho^{Bag}=\Omega+B+\sum_{i}\mu_{i} n_{i}, \label{Eq26}
\end{equation}
In Eq.~(\ref{Eq25}), the last term in the right hand side arises from the consideration that $B$ depends on energy density. The additional term vanishes for constant $B$ and subsequently the EoS of the interior matter reduces to that obtained in MIT bag model.

\section{EOS in the CFL phase}\label{eos}
It is widely accepted that if the strange quark mass $m_{s}$ is small enough, the CFL state would be the minimum energy configuration at high densities. In case of CFL phase of SQM, the constituent $u$, $d$ and $s$ quarks together form pairs and are compelled to pick up equal Fermi momenta \cite{Zhu}. To obtain the EoS and relevant quantities one has to start with the thermodynamic potential density of the SQM in the context of CFL phase. In the CFL phase of SQM at $T\rightarrow 0$, the thermodynamic potential density is given by \cite{Prasad,Alford1,Alford2}.
\begin{eqnarray}
\Omega&=&\frac{6}{\pi}\int_{0}^{\nu}(p-\mu)p^{2}dp-\frac{3}{\pi^{2}}\Delta^{2}\mu^{2} \nonumber \\ 
&& +\frac{3}{\pi^{2}}\int_{0}^{\nu}[{(p^{2}+m_{s}^{2})}^{\frac{1}{2}}-\mu]p^{2}dp, \label{Eq28}
\end{eqnarray}
where $\mu=\frac{\mu_{u}+\mu_{d}+\mu_{s}}{3}$, $\mu_{u}$, $\mu_{d}$ and $\mu_{s}$ represent respectively chemical potential of the quarks $u$, $d$ and $s$ and $m_{s}$ is the strange quark mass. $\Delta$ represents the gap parameter in color-superconductivity. In the CFL phase Fermi momentum of $u$, $d$ and $s$ quarks are same \cite{Prasad,Zhu}. This common Fermi momentum is determined by minimizing the free energy density up to order $m_{s}^{4}$ and is given by
\begin{equation}
\nu=2\mu-{(\mu^{2}+\frac{m_{s}^{2}}{3})}^{\frac{1}{2}}=\mu-\frac{m_{s}^{2}}{6\mu}+\frac{m_{s}^{4}}{72\mu^{3}}. \label{Eq29}
\end{equation}
Substituting Eq.~(\ref{Eq28}) in Eqs.~(\ref{Eq21})-(\ref{Eq23}), we get the EOS in the CFL phase.
\begin{equation}
p=\frac{3\mu^{4}}{4\pi^{2}}-\frac{3m_{s}^{2}\mu^{2}}{4\pi^{2}}+\frac{1-12ln(\frac{m_{s}}{2\mu})}{32\pi^{2}}m_{s}^{4}+\frac{3}{\pi^{2}}\Delta^{2}\mu^{2}-B+n_{b}\frac{\partial B}{\partial n_b}, \label{Eq30}
\end{equation}
\begin{equation}
\rho=\frac{9\mu^{4}}{4\pi^{2}}-\frac{3m_{s}^{2}\mu^{2}}{4\pi^{2}}+\frac{11m_{s}^{4}}{32\pi^{2}}+\frac{3ln(\frac{m_{s}}{2\mu})}{8\pi^{2}}m_{s}^{4}+\frac{3}{\pi^{2}}\Delta^{2}\mu^{2}+B, \label{Eq31}
\end{equation}
\begin{equation}
n=\frac{\nu^{3}+2\Delta^{2}\mu}{\pi^{2}}=\frac{\mu^{3}}{\pi^{2}}-\frac{m_{s}^{2}\mu}{2\pi^{2}}+\frac{m_{s}^{4}}{8\pi^{2}\mu}+\frac{2\Delta^{2}\mu}{\pi^{2}}. \label{Eq32}
\end{equation}

\begin{figure}[ht!]
\centering
\includegraphics[width=8cm]{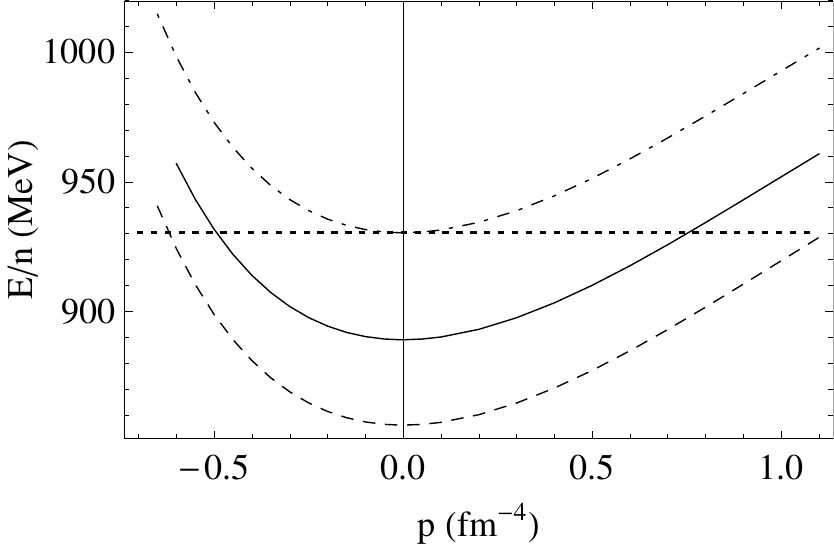}
\caption{Energy per baryon ($\frac{E}{n}$) vs pressure $p$ of SQM in the CFL phase. Here dashed line is drawn for $m_{s}=50~MeV$, solid line and dotdashed line represent $m_{s}=150~MeV$ and $m_{s}=228.3~MeV$ respectively. The energy per baryon for the horizontal line is 930.4 $MeV$ which is the typical energy per baryon of $^{56}Fe$. To draw the plots we have taken $\Delta=100~MeV$.}
\label{Fig2}
\end{figure}

\begin{figure}[ht!]
\centering
\includegraphics[width=8cm]{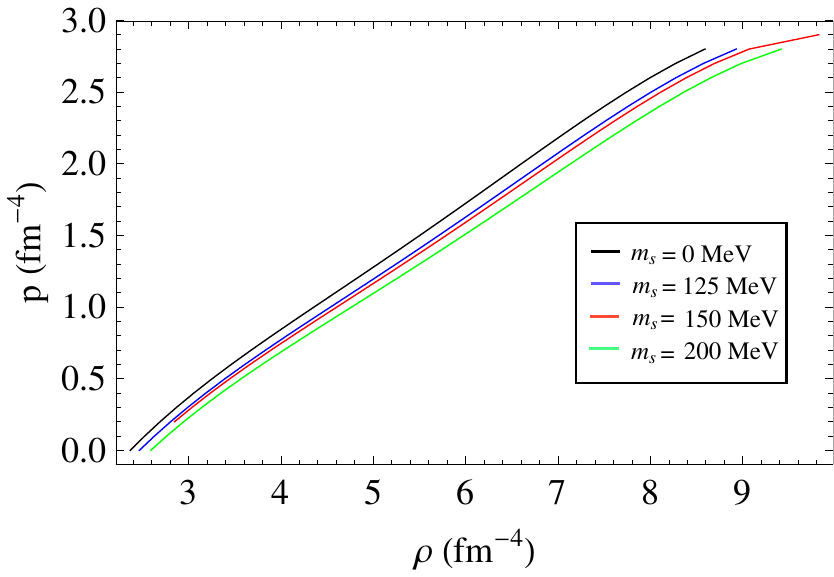}
\caption{Equation of state of SQM in the CFL phase with density dependent $B$.}
\label{Fig3}
\end{figure}
Zhu {\it et al.} \cite{Zhu} considered density dependent $B$ given in Eq.~(\ref{Eq10}) predicted by Aguirre \cite{Aguirre}. Using this density dependence of $B$, the energy associated to each baryon ($E/n$) is plotted as a function of pressure $p$ and is shown in Fig.~\ref{Fig2}. In this model zero pressure appears exactly at the lowest energy state. The minimum energy per baryon for $m_s=228.3~MeV$ touches the line for $^{56}Fe$. Therefore as per stability of 3-flavour quarks is concerned the upper limit of $m_s$ is 228.3 $MeV$. The EoS in this model has been visualized in Fig.~\ref{Fig3}.

\section{Vaidya-Tikekar model EoS of different compact stars}\label{vaidya}
In the standard procedure, to get the stellar structure of compact star one has to solve the TOV equation for given EoS, using appropriate boundary conditions. However, in Vaidya-Tikekar model a geometry is given and then one looks for the suitable composition of matter to support this geometry. The geometry is described by one free parameter $\lambda$ known as spheroidal parameter and other is known as curvature parameter $R$ to be determined. The EoS obtained from V-T model picks up linear form when the value of the parameter $\lambda$ is large. In this geometrical model the stellar structures are found to be stable against small radial oscillations. The EoS predicted by Gondek-Rosi{\'{n}}ska {\it et al.} \cite{Gondek} is in well agreement with the EoS predicted by Sharma {\it et al.} \cite{Sharma2} for SAX J 1808.4-3658 from the V-T geometrical model. Thus it is interesting to note that the EoS obtained by considering geometry attached to physical space of the compact object is consistent with that obtained from the microscopic composition of the interior matter of compact object as both of them give stable compact object having same mass and radius. In VT model \cite{Vaidya} a specific form of $g_{rr}$ metric component is prescribed and using this metric ansatz the general solution for the other metric component $g_{tt}$ is obtained following the process prescribed by Mukherjee {\it et al.} \cite{Mukherjee}. The general solution is briefly outlined below.\\
We consider a static, spherically symmetric star whose interior metric is given by
\begin{equation}
ds^2=-e^{2\gamma(r)}dt^2+e^{2\xi(r)}dr^2+r^2(d{\theta}^2+sin^2{\theta} d{\phi}^2). \label{Eq33}
\end{equation}
Considering the ansatz, predicted by Vaidya and Tikekar \cite{Vaidya}, given below
\begin{equation}
e^{2\xi}=\frac{1+\lambda \frac{r^2}{R^2}}{1-\frac{r^2}{R^2}}, \label{Eq34}
\end{equation}
and assuming that inside the star matter distribution is like perfect fluid with anisotropy in pressure, we use the approach of Goswami {\it et al.} \cite{Goswami} which is the anisotropic extension of the solution obtained by Mukherjee {\it et al.} \cite{Mukherjee} and obtain the solution for the $g_{tt}$ component of metric function,
\begin{equation}
\psi(z)=e^{\gamma}=A\left[\frac{cos[(n+1)\zeta +\delta]}{n+1}-\frac{cos[(n-1)\zeta +\delta]}{n-1}\right], \label{Eq35}
\end{equation}
where $\zeta=cos^{-1}z$, $z^{2}=\frac{\lambda}{\lambda+1}(1-\frac{r^2}{R^2})$ and $n^{2}=\lambda(1-\alpha)+2$. The energy-density $\rho$,radial ($p_r$) and transverse pressures ($p_t$) in this model are given by
\begin{equation}
\rho=\frac{1}{R^{2}(1-z^{2})}\left[1+\frac{2}{(\lambda+1)(1-z^{2})}\right], \label{Eq36}
\end{equation}
\begin{equation}
p_r=-\frac{1}{R^{2}(1-z^{2})}\left[1+\frac{2z}{(\lambda+1)}\left(\frac{\psi_{z}}{\psi}\right)\right], \label{Eq37}
\end{equation}
\begin{equation}
p_t=p_r+\Delta, \label{Eq38}
\end{equation}
where $\psi_{z}$ represents the first derivative of function $\psi$ with respect to variable $z$ and $\Delta=\frac{\alpha \lambda[\lambda-(\lambda+1)z^2]}{R^2(\lambda+1)^2(1-z^2)^2}$ is the measure of the pressure anisotropy parametrized by $\alpha$. The expression for the total mass contained within the radius $b$ of a star is given below:
\begin{equation}
M(b)=\frac{(1+\lambda)\frac{b^3}{R^2}}{2(1+\lambda \frac{b^2}{R^2})}. \label{Eq39}
\end{equation}
The model have four parameters, $A$, $\delta$, $R$ and $\lambda$. Where $\lambda$ is a input parameter and the rest will be fixed by imposing restrictions at the boundary of a star as given below.
\begin{enumerate}
\item At the boundary ($r=b$) of a star the interior metric should be finite and matched with the value obtained from Schwarzschild exterior metric given below:
\begin{eqnarray}
ds^2&=&-\left(1-\frac{2M}{r}\right)dt^2+\left(1-\frac{2M}{r}\right)^{-1}dr^2 \nonumber \\
&&+r^2\left(d\theta^2+sin^2\theta d\phi^2\right). \label{Eq40}
\end{eqnarray}
Therefore matching of the metric (\ref{Eq33}) and (\ref{Eq40}) at the boundary yields
\begin{equation}
e^{2\xi}=\frac{1+\lambda \frac{b^2}{R^2}}{1-\frac{b^2}{R^2}}=\left(1-\frac{2M}{b}\right)^{-1}. \label{Eq41} 
\end{equation}
\begin{equation}
e^{2\gamma}=\left(1-\frac{2M}{b}\right). \label{Eq42}
\end{equation}
\item Again at the surface of a star the radial pressure may be equated to zero which gives the following condition
\begin{equation}
\left(\frac{\psi_z}{\psi}\right)_{z_b}=-\frac{(\lambda+1)}{2z_b}. \label{Eq43}
\end{equation}
Again taking derivative of $\psi$ in Eq.~(\ref{Eq35}) with respect to $z$ we get the ratio at $z=z_b$
\begin{equation}
\resizebox{0.48\textwidth}{!}{$\left(\frac{\psi_{z}}{\psi}\right)_{z_b}=\frac{n^2-1}{\sqrt{1-z_b^2}}\left[\frac{sin[(n+1)\zeta+\delta]-sin[(n-1)\zeta+\delta]}{(n-1)cos[(n+1)\zeta+\delta]-(n+1)cos[(n-1)\zeta+\delta]}\right]$}. \label{Eq44}
\end{equation}
\end{enumerate}
Thus equating Eqs.~(\ref{Eq43}) and ~(\ref{Eq44}) we can obtain the value of $\delta$. As in this model $\lambda$ specifies the EoS of the interior matter of a star of given mass ($M$) and radius ($b$), choice of any one among radius ($b$), surface density ($\rho_b$) or central density ($\rho_0$) will determine the value of $R$ for a given choice of $\lambda$. Thus the present model accounts for a complete description of physical properties of the star which includes the radial variation of energy density and radial pressure using Eqs.~(\ref{Eq36}) and ~(\ref{Eq37}). The value of $\rho$ and $p_r$ then used to determine the EoS of the interior matter. 

\section{Comparison of thermodynamic model with exactly solvable model}\label{physical}
\begin{figure}[ht!]
\centering
\includegraphics[width=8cm]{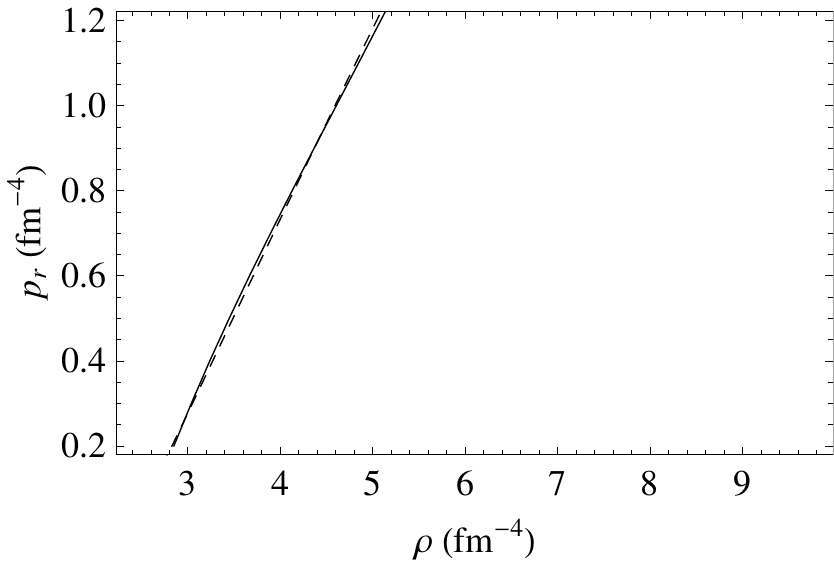}
\caption{Equation of state of SQM in the CFL phase with density dependent $B$ (solid line) obtained from the thermodynamic model for $m_s=150$ $MeV$ and those obtained from VT model for the compact star 4U 1820-30 (dashed line) with predicted radius $b=8.54$ $km$. The model parameters are $\lambda=5$ and $\alpha=0$}
\label{Fig4}
\end{figure}

\begin{figure}[ht!]
\centering
\includegraphics[width=8cm]{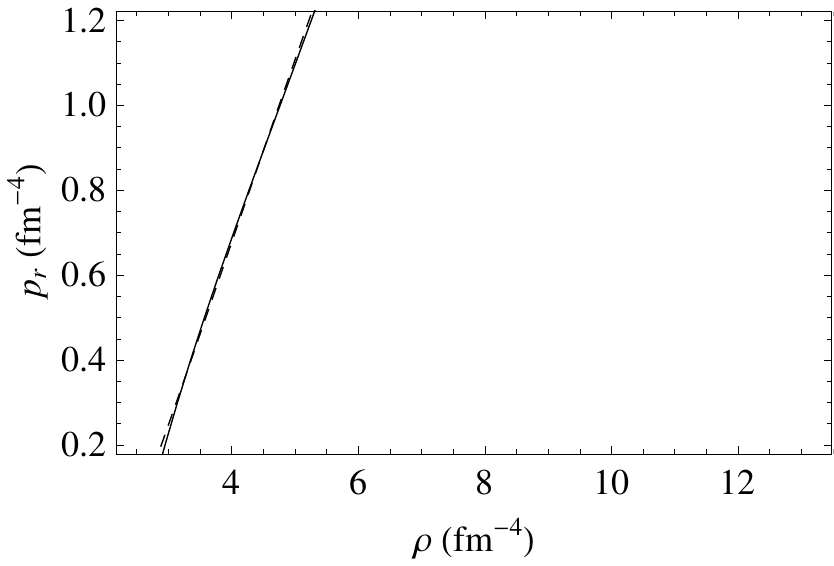}
\caption{Equation of state of SQM in the CFL phase with density dependent $B$ (solid line) obtained from the thermodynamic model for $m_s=200$ $MeV$ and those obtained from VT model for the compact star PSR J1614-2230 (dashed line) with predicted radius $b=8.46$ $km$. The model parameters are $\lambda=50$ and $\alpha=0.39$}
\label{Fig5}
\end{figure}

\begin{figure}[ht!]
\centering
\includegraphics[width=8cm]{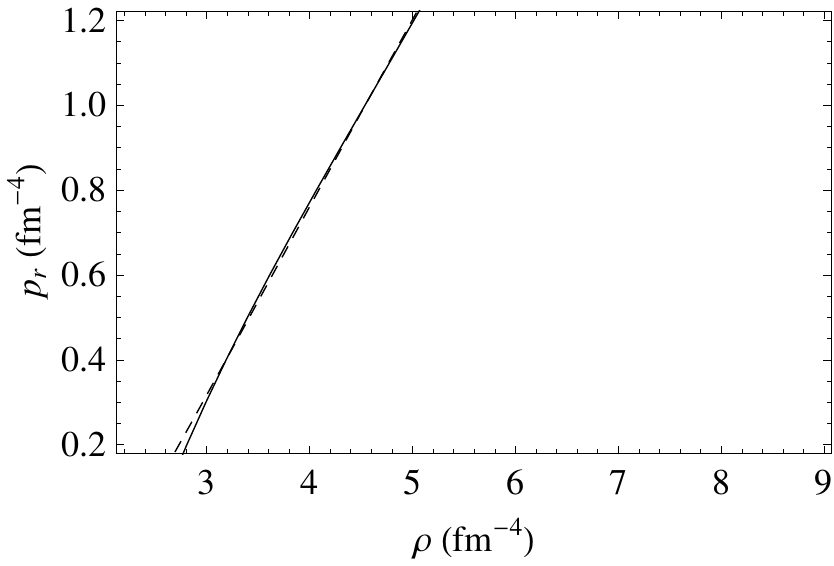}
\caption{Equation of state of SQM in the CFL phase with density dependent $B$ (solid line) obtained from the thermodynamic model for $m_s=125$ $MeV$ and those obtained from VT model for the compact star PSR J1903+0327 (dashed line) with predicted radius $b=8.5$ $km$. The model parameters are $\lambda=100$ and $\alpha=0$}
\label{Fig6}
\end{figure}

As we already have obtained the expressions for physical parameters like energy density and pressures, we are now going to apply our model to some strange star candidates such as 4U 1820-30 \cite{Guver}, PSR J1614-2230 \cite{Miller}, PSR J1903+0327 \cite{Freire}, PSR J0030+0451 \cite{Miller2} and PSR J0740+6620 \cite{Riley}. In Vaidya-Tikekar model the EoS of a star is dependent on both $\lambda$ and $\alpha$, therefore by changing any of them we can match the EoS of a given star with that obtained from thermodynamic point of view. In the present set up we note that to fit the EoS of a star with the CFL EoS the radius may be predicted. In Figs.~\ref{Fig4} - \ref{Fig6} we have plotted the EoS of the chosen compact objects and compared with that of CFL state. It is very interesting to note that our model permits a wide range of the model parameters $\lambda$ and $\alpha$ for which strange quarks having CFL type EoS may be perceived inside compact objects. Another noteworthy feature is that to have a CFL type EoS, PSR J1614-2230 should be anisotropic in nature. In Table~\ref{tab1} we have tabulated the predicted radius and central density of compact objects as mentioned above by using (i) CFL EoS and (ii) MIT bag EoS. In CFL phase due to cooper pair formation, quarks collectively behave as bosons thereby overlooks Pauli exclusion principle. Which admits a more closed pack structure and consequently central density of star increases, gravity on the other hand finds a new path to squeeze the star to a more smaller size. This can be noted from Table~\ref{tab1}.  

\begin{table*}
\flushleft
\caption{Tabulation of predicted radius and central density of some compact objects obtained from (i) CFL Eos and (ii) MIT bag Eos.}
\label{tab1}
\begin{tabular*}{\columnwidth}{@{\extracolsep{\fill}}ccccccccc@{}}
\cline{1-9}
Compact & \multirow{2}{*}{Mass ($M_{\odot}$)} & \multirow{2}{*}{$\lambda$} & \multirow{2}{*}{$\alpha$} & \multirow{2}{*}{$m_s$ ($MeV$)} & \multicolumn{2}{c}{predicted radius ($km$)} & \multicolumn{2}{c}{central density $\times$$10^{15}$ ($g/cm^3$)} \\ \cline{6-9}
Object  &   &   &      &          &  CFL EoS  &  MIT bag EoS  & CFL Eos  & MIT bag EoS \\ \cline{1-9}
4U 1820-30 & $1.58^{+0.06}_{-0.06}$ &5 & 0& 150 & 8.54 & 10.85 & 2.20 & 0.91 \\   
PSR J1614-2230& $1.928^{+0.017}_{-0.017}$ &50&0.39& 200& 8.46 & 8.75 & 4.42 & 3.75 \\
PSR J1903+0327& $1.667^{+0.021}_{-0.021}$ &100& 0& 125 &  8.50 & 9.85 & 3.00 & 1.63 \\ 
PSR J0030+0451& $1.44^{+0.15}_{-0.14}$ &2.2& 0& 100& 8.70 & 13 & 1.55 & 0.40 \\      
PSR J0740+6620& $2.072^{+0.067}_{-0.066}$ &100& 0.45& 100& 8.7 & 9.07 & 4.88 & 3.94 \\
\cline{1-9}                    
\end{tabular*}
\end{table*}

\section{Mass-Radius relation of strange star admitting CFL EoS}\label{m-r}
\begin{figure}[ht!]
\centering
\includegraphics[width=8.5cm]{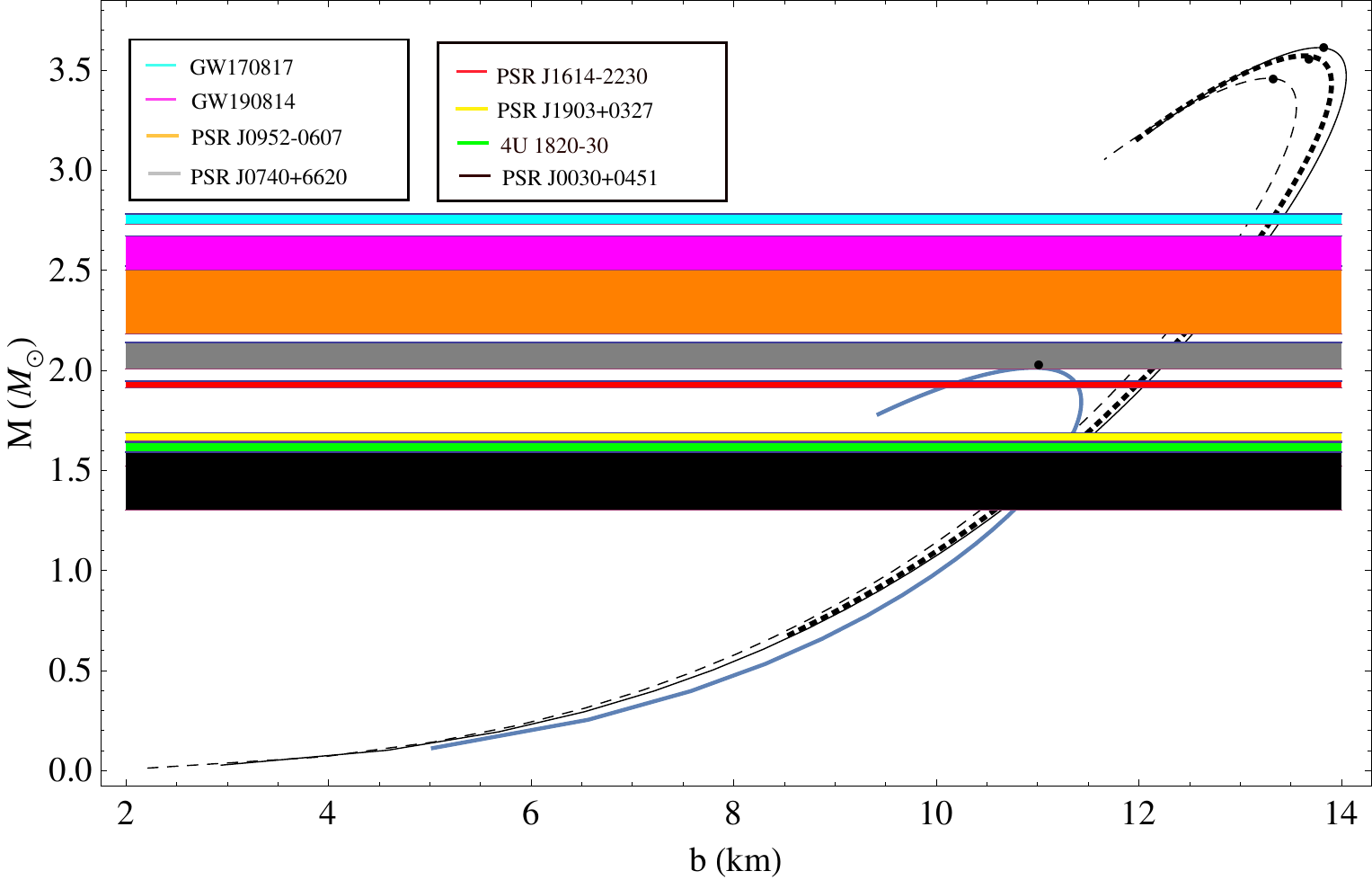}
\caption{Mass - radius plot of strange star composed of strange quark matter described by CFL EoS. Here solid, dotted and dashed line correspond to $m_s=0$, 100 and 150 $MeV$ respectively. Blue curve shows the mass radius relation for quark matter following the MIT bag EoS $p_r=\frac{1}{3}(\rho-4B)$ with $B=57.55$ $MeV/fm^3$. Mass ranges of different compact objects are also shown and indicated in the figure.}
\label{Fig7}
\end{figure}

We now study the mass - radius curve of compact objects by solving TOV equation comprised of strange matter having CFL type EoS. The mass - radius plot is shown in Fig.~\ref{Fig7}.

\begin{figure}[ht!]
\centering
\includegraphics[width=8.5cm]{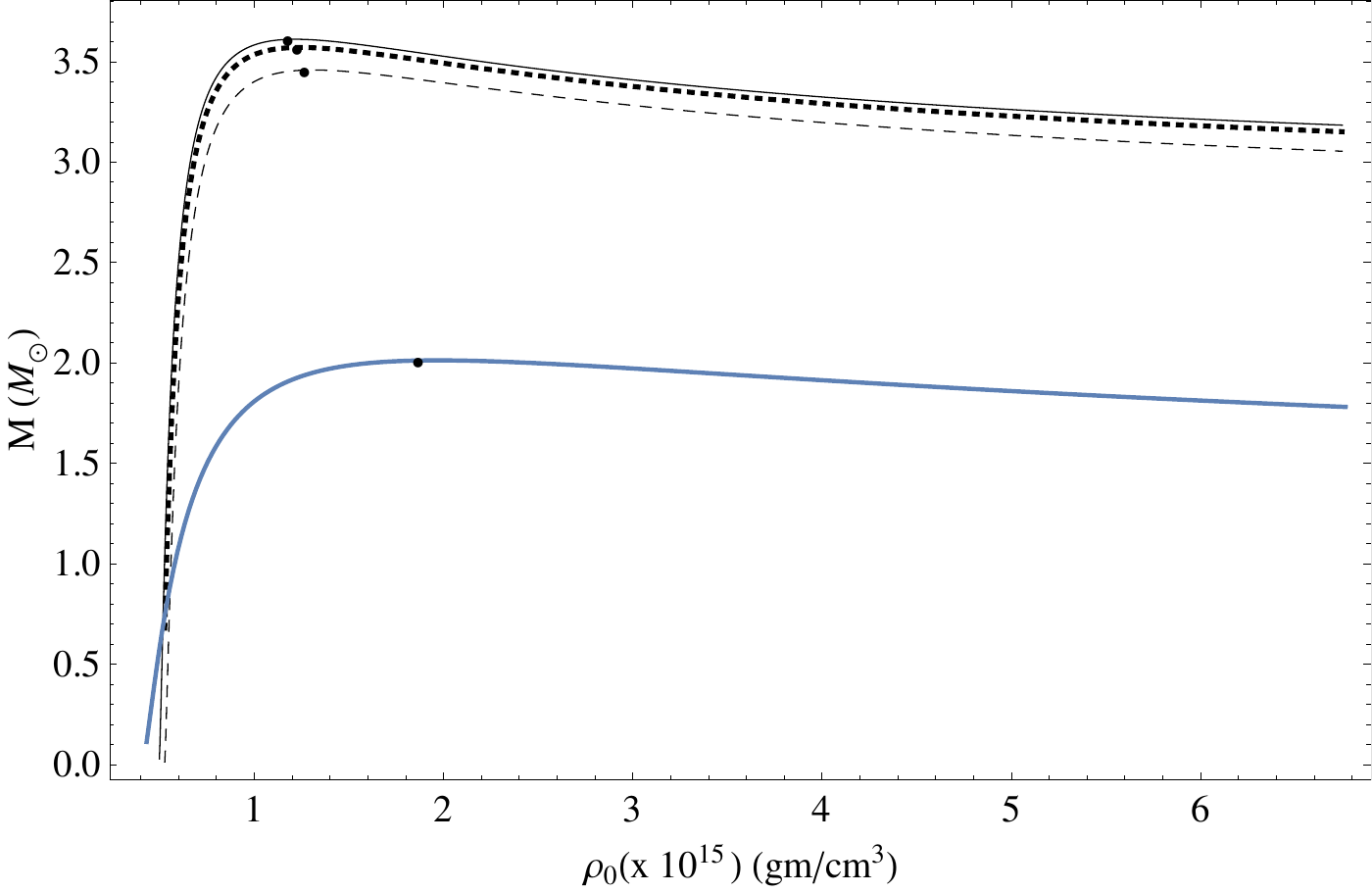}
\caption{Mass vs central density plot of strange star composed of strange quark matter described by CFL EoS. Here solid, dotted and dashed line correspond to $m_s=0$, 100 and 150 $MeV$ respectively. Blue curve shows the mass radius relation for quark matter following the MIT bag EoS $p_r=\frac{1}{3}(\rho-4B)$ with $B=57.55$ $MeV/fm^3$}
\label{Fig8}
\end{figure}
From mass-radius plot the maximum mass of strange star in CFL EoS is found to be maximum for $m_s=0$ $MeV$ which is $\sim 3.61$ $M_\odot$ and decreases when $m_s$ increases. As for instance maximum mass is 3.570 $M_\odot$ when $m_s= 100$ $MeV$ while it reduces to the value 3.458 $M_\odot$ when $m_s= 150$ $MeV$. These values of maximum masses are well above the value 2.03 $M_\odot$ obtained by considering MIT bag EoS for massless quarks ($m_s=0$ $MeV$). Thus we may conclude that CFL EoS admits wider range of compact objects as indicated in the mass-radius plot in Fig.~\ref{Fig7}. The mass - central density plot reveals that maximum density obtained from this model is 1.221 $\times 10^{15}$ $gm/cm^3$ for $m_s = 0$ $MeV$ while for other values of $m_s$ it takes lower value. Above the maximum point mass ($M$) of a compact object decreases with an increase in the value of central density ($\rho_0$) i.e. $(\frac{\partial{M}}{\partial{\rho_0}}<0)$ which corresponds to a collapsible state \cite{Zeldovich} and is not allowed as per stability is concerned. Interestingly the maximum mass takes higher value in CFL phase than MIT bag model as can be observed from Fig.~\ref{Fig7}. A physical justification can be made as follows: from Fig.~\ref{Fig8} it is noted that maximum central density of a star takes lower value in CFL phase than in MIT bag model EoS. Therefore as quarks form cooper pairs in colour superconducting state it takes away some part of the stars energy which is balanced by incorporating extra mass.   
\section{Energy Conditions}\label{ec}
For a relativistic sphere a physically realistic model is possible if the following energy conditions are satisfied at all internal points and surface of the sphere \cite{Brassel1,Brassel2}. 

\begin{enumerate}
\item Null Energy Condition (NEC) : $\rho + p_{r}\geq 0$; $\rho + p_{t} \geq 0$. 
\item Weak Energy Condition (WEC) : $\rho + p_{r}\geq 0$; $\rho \geq 0, \rho + p_{t} \geq 0$. 
\item Strong Energy Condition (SEC) : $\rho + p_{r}\geq 0$; $\rho + p_{r} + 2p_{t} \geq 0$. 
\item Dominant Energy Condition (DEC) : $\rho \geq 0$; $\rho - p_{r} \geq 0$; $\rho - p_{t} \geq 0$. 
\end{enumerate}

\begin{figure}[ht!]
\begin{center}
\includegraphics[width=8cm]{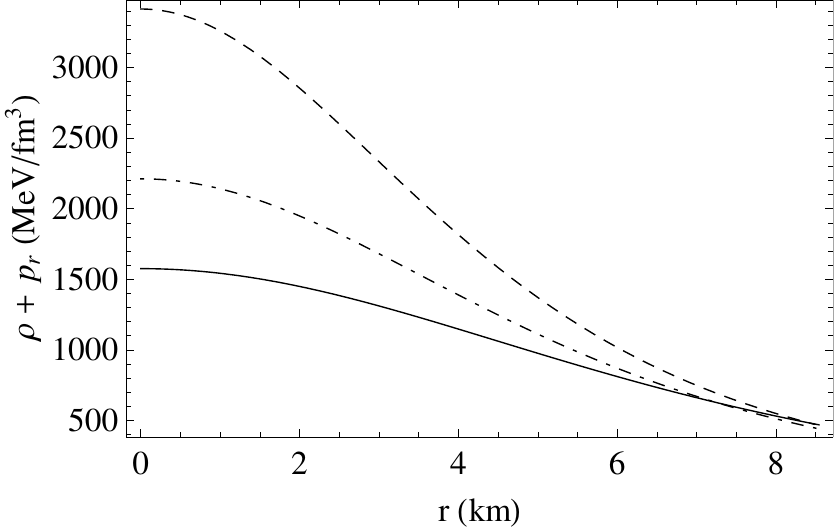}
\caption{Variation of ($\rho+p_{r}$) with radial distance $r$ for different compact objects. Here solid, dotdashed and dashed lines are drawn for 4U 1820-30, PSR J1903+327 and PSR J1614-2230 respectively}
\label{Fig9}
\end{center}
\end{figure}

\begin{figure}[ht!]
\begin{center}
\includegraphics[width=8cm]{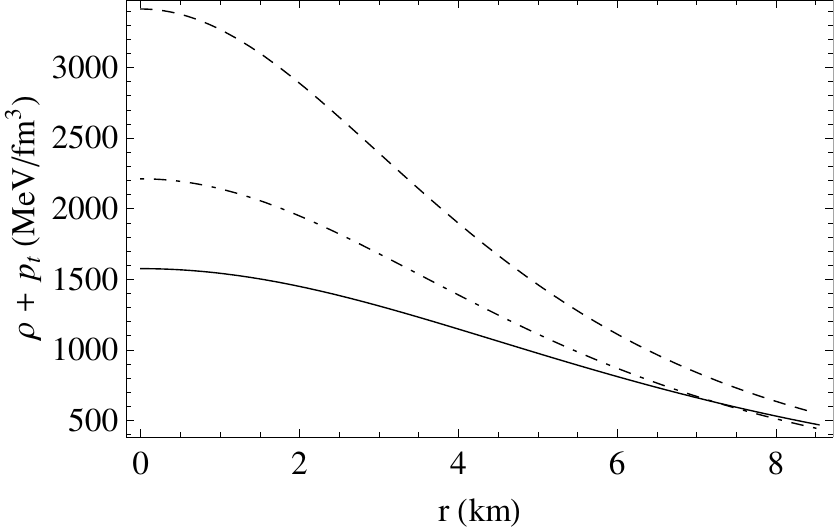}
\caption{Variation of ($\rho+ p_{t}$) with radial distance $r$ for different compact objects. Here solid, dotdashed and dashed lines are drawn for 4U 1820-30, PSR J1903+327 and PSR J1614-2230 respectively}
\label{Fig10}
\end{center}
\end{figure}

\begin{figure}[ht!]
\begin{center}
\includegraphics[width=8cm]{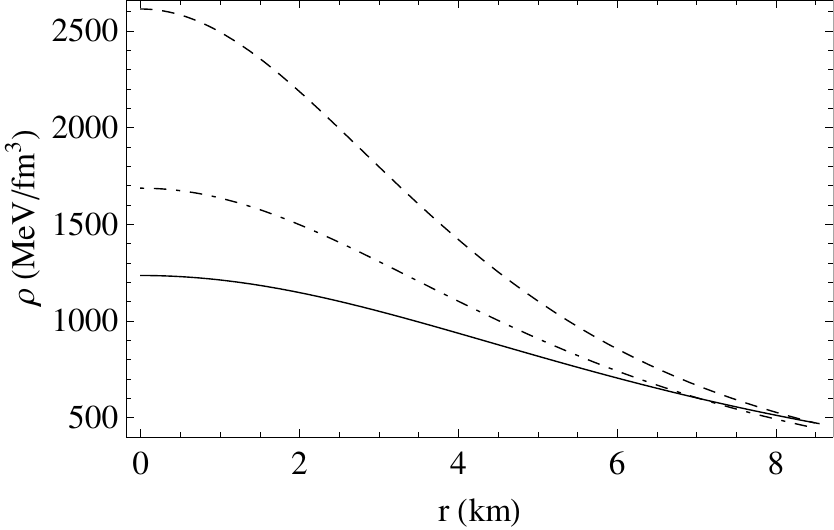}
\caption{Variation of $\rho$ with radial diatance $r$ for different compact objects. Here solid, dotdashed and dashed lines are drawn for 4U 1820-30, PSR J1903+327 and PSR J1614-2230 respectively}
\label{Fig11}
\end{center}
\end{figure}

\begin{figure}[ht!]
\begin{center}
\includegraphics[width=8cm]{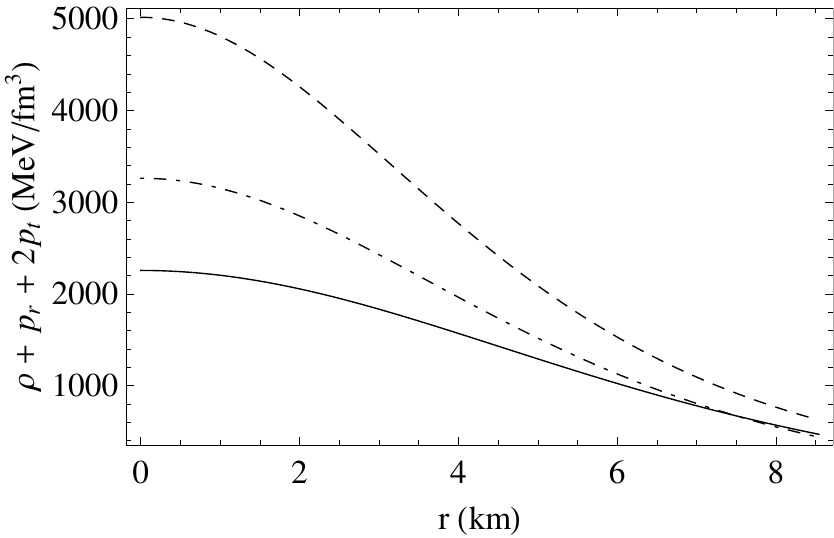}
\caption{Variation of ($\rho + p_{r} + 2p_{t}$) with radial distance $r$ for different compact objects. Here solid, dotdashed and dashed lines are drawn for 4U 1820-30, PSR J1903+327 and PSR J1614-2230 respectively}
\label{Fig12}
\end{center}
\end{figure}

\begin{figure}[ht!]
\begin{center}
\includegraphics[width=8cm]{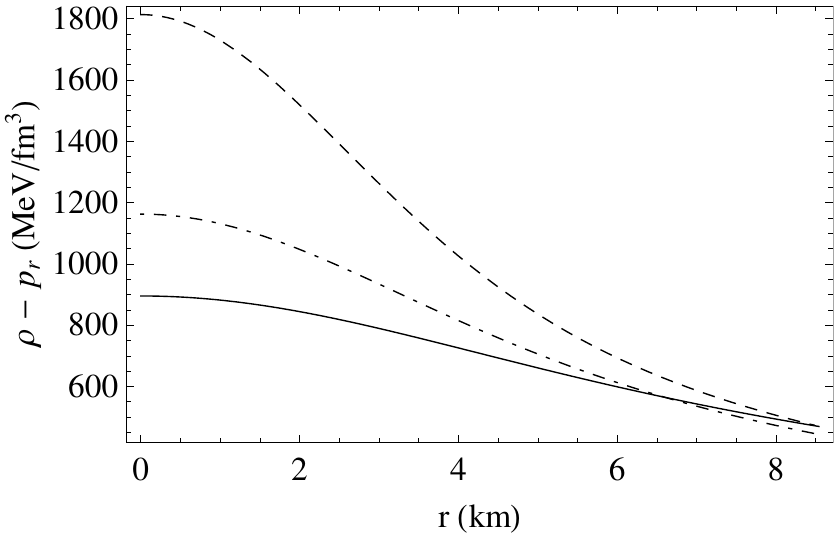}
\caption{Variation of ($\rho- p_{r}$) with radial distance $r$ for different compact objects. Here solid, dotdashed and dashed lines are drawn for 4U 1820-30, PSR J1903+327 and PSR J1614-2230 respectively}
\label{Fig13}
\end{center}
\end{figure}

\begin{figure}[ht!]
\begin{center}
\includegraphics[width=8cm]{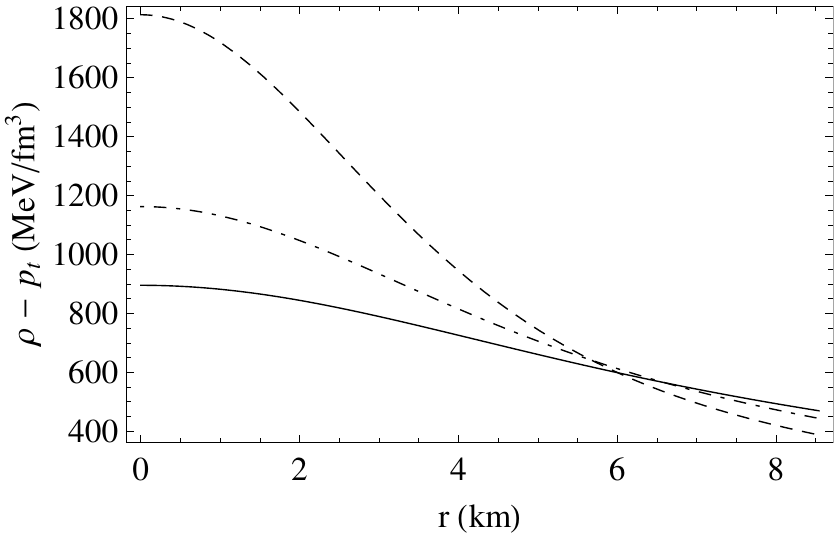}
\caption{Variation of ($\rho- p_{t}$) with radial distance $r$ for different compact objects. Here solid, dotdashed and dashed lines are drawn for 4U 1820-30, PSR J1903+327 and PSR J1614-2230 respectively}
\label{Fig14}
\end{center}
\end{figure}

In Figs.~\ref{Fig9} and \ref{Fig10}, we have plotted the Null Energy Conditions. In Figs. ~\ref{Fig9}, \ref{Fig10} and ~\ref{Fig11}, we have shown the Weak Energy Conditions. In Figs. ~\ref{Fig9} and ~\ref{Fig12}, Strong Energy Conditions have been visualized. Figs.~\ref{Fig13} and \ref{Fig14} depict the Dominant Energy Conditions. From Figs.~\ref{Fig9} - \ref{Fig14}, we note that in this model, necessary energy conditions held good.  

\section{Equilibrium under different forces}\label{eq}  
\begin{figure}[ht!]
\begin{center}
\includegraphics[width=9cm]{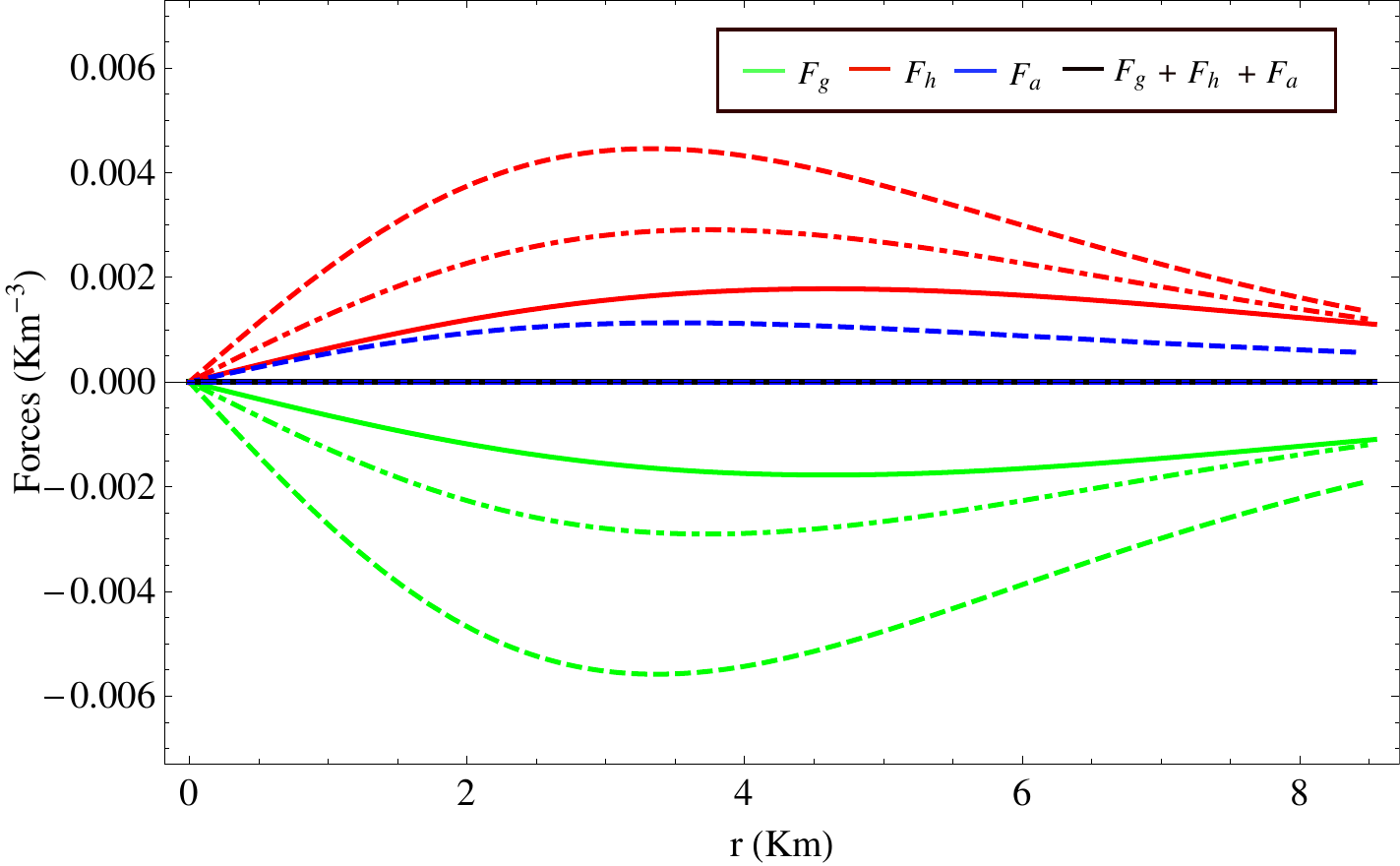}
\caption{Variation of different forces with radial distance $r$ inside various compact objects. Here solid, dotdashed and dashed lines are drawn for 4U 1820-30, PSR J1903+0327 and PSR J1614-2230 respectively.}
\label{Fig15}
\end{center}
\end{figure}

The equilibrium of an anisotropic fluid sphere is described the generalized TOV equation. Which has the following form as presented by Ponce de Le{\'o}n \cite{Ponce}
\begin{equation}
-\frac{M_G (\rho+p_r)}{r^2}e^{(\xi-\gamma)} -\frac{dp_r}{dr} + \frac{2}{r}(p_t-p_r) = 0, \label{Eq45}
\end{equation}
in the above equation $M_G$ denotes the active gravitational mass contained within a sphere of radius $r$ and can be obtained from the formula given by Tolman-Whittaker and Einstein field equations having the following form
\begin{equation}
M_G(r) = \gamma^{\prime}r^2 e^{(\gamma-\xi)}. \label{Eq46}
\end{equation}
Using Eq.~(\ref{Eq46}), Eq.~(\ref{Eq45}) may be recasted in the form
\begin{equation}
F_g + F_h + F_a = 0, \label{Eq47}
\end{equation}
where $F_g=-\gamma^{\prime}(\rho+p_r)$, $F_h=-\frac{dp_r}{dr}$ and $F_a=\frac{2\Delta}{r}$. The condition of equilibrium of an anisotropic distribution of matter under the mutual influence of force due to gravity ($F_g$), force due to anisotropy ($F_a$) and hydrostatic force ($F_h$) is represented by Eq.~(\ref{Eq47}). The nature of these forces are depicted in Fig.~\ref{Fig15}. It may be noted that all of these forces increase from centre attain maximum value at some interior point then decrese upto surface. The sum of all forces vanishes at all interior points as indicated in black line. Surprisingly the anisotropic force is only relevant for PSR J1614-2230 since it shows anisotropic behaviour to obey the CFL phase EoS whereas 4U 1820-30 and PSR J1903+0327 might be isotropic to have a CFL like EoS for its interior matter. Thus our model permits equilibrium of various compact objects in presence of anisotropic pressure.

\section{Herrera cracking condition}\label{hr}
\begin{figure}[ht!]
\begin{center}
\includegraphics[width=8cm]{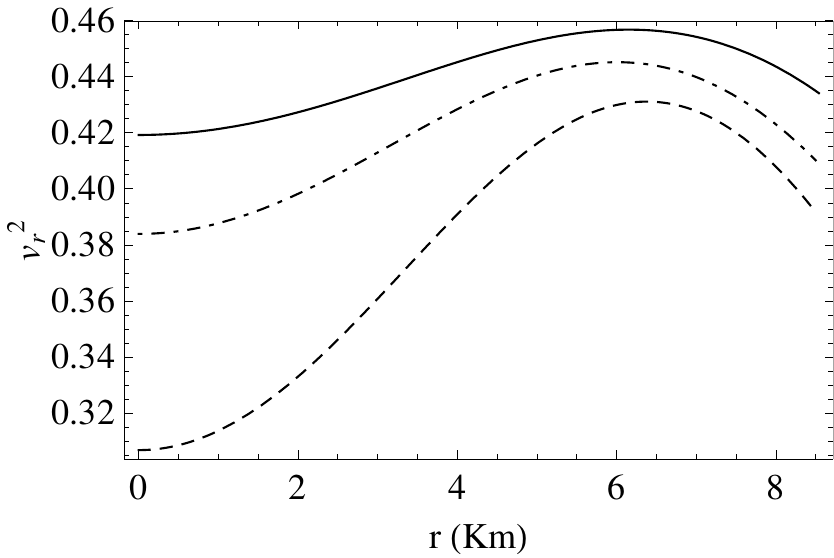}
\caption{Variation of $v_r^2$ with radial distance $r$ inside various compact objects. Here solid, dotdashed and dashed lines are drawn for 4U 1820-30, PSR J1903+0327 and PSR J1614-2230 respectively.}
\label{Fig16}
\end{center}
\end{figure}

\begin{figure}[ht!]
\begin{center}
\includegraphics[width=8cm]{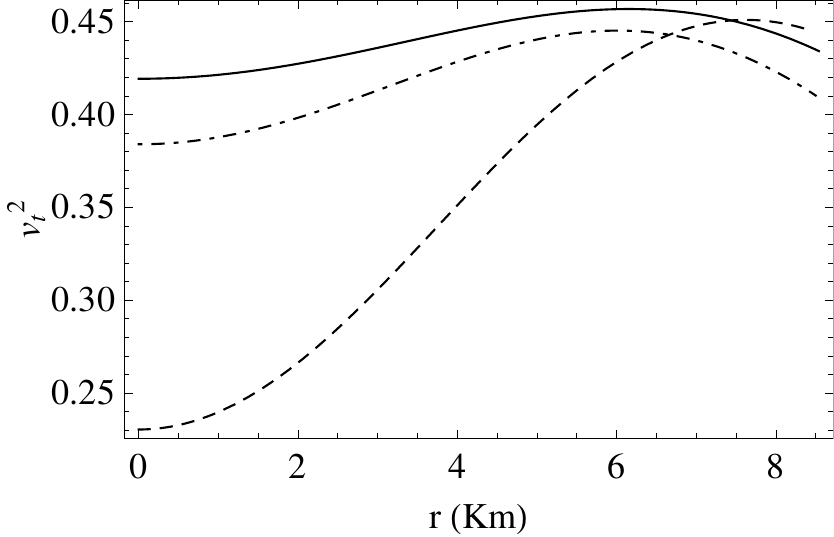}
\caption{Variation of $v_t^2$ with radial distance $r$ inside various compact objects. Here solid, dotdashed and dashed lines are drawn for 4U 1820-30, PSR J1903+0327 and PSR J1614-2230 respectively.}
\label{Fig17}
\end{center}
\end{figure}

We know that a stellar model can be physically acceptable if both the square of radial ($v_r^2=\frac{dp_r}{d\rho}$) and transeverse ($v_t^2=\frac{dp_t}{d\rho}$) sound velocities remain below the velocity of light inside a star which is referred to as 'causality condition'. In Figs.~\ref{Fig16} and \ref{Fig17}, we have plotted variation of $v_r^2$ and $v_t^2$ with $r$ for different compact objects. It is found that causality condition is obeyed in this model.

\begin{figure}[ht!]
\begin{center}
\includegraphics[width=8cm]{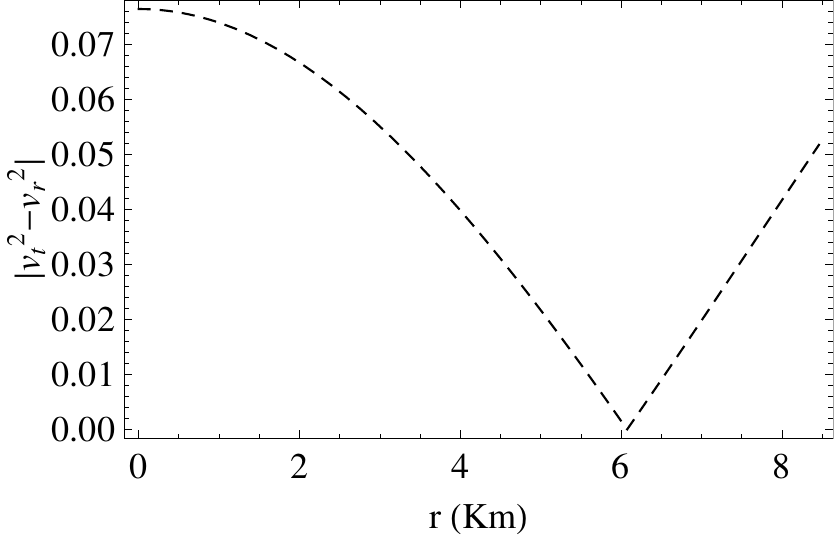}
\caption{Variation of $|v_t^2-v_r^2|$ with radial distance $r$ inside PSR J1614-2230.}
\label{Fig18}
\end{center}
\end{figure}

To check the stability of a stellar system the concept 'cracking' was introduced by Herrera \cite{Herrera}. Based on this method earlier Abreu {\it et al.} \cite{Abreu} gave a criteria 
\begin{equation}
0 \le |v_{t}^{2}-v_{r}^2| \le 1. \label{Eq48}
\end{equation} 
It is to be noted that for isotropic star $\frac{dp_r}{d\rho}$ $=\frac{dp_t}{d\rho}$ and therefore for 4U 1608-52 and PSR J1903+0327 we must have $|v_{t}^{2}-v_{r}^2|=0$ at all interior points. Therefore we have checked the validity of Abreu's inequality for PSR J1614-2230 only which is shown in Fig.~\ref{Fig18}

\begin{figure}[ht!]
\begin{center}
\includegraphics[width=8cm]{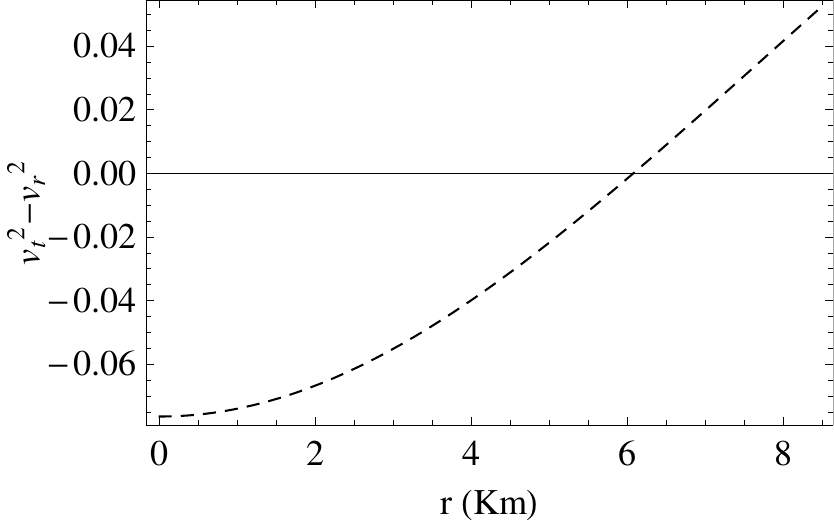}
\caption{Variation of $(v_t^2-v_r^2)$ is shown against radial distance $r$ inside PSR J1614-2230.}
\label{Fig19}
\end{center}
\end{figure}

Again the condition $-1\leq v_t^2-v_r^2 \leq 1$ implies two disctinct regions
\begin{eqnarray}
-1\leq (v_t^2-v_r^2) \leq 0 , ~\text{potentially stable}, \nonumber \\
0\leq (v_t^2-v_r^2) \leq 1 , ~\text{potentially unstable}. \nonumber
\end{eqnarray}
The plot in Fig.~\ref{Fig19} indicates that there is region inside the star where $(v_t^2-v_r^2)$ alters sign and therefore changes from a potentially stable to a potentially unstable region.

\section{Adiabatic index}\label{ad}
\begin{figure}[ht!]
\begin{center}
\includegraphics[width=8cm]{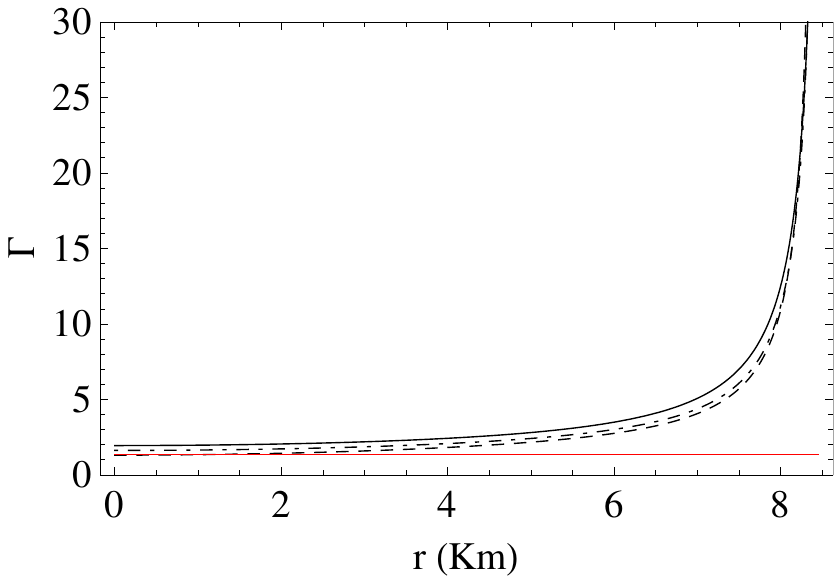}
\caption{Variation of adiabatic index ($\Gamma$) against radial distance $r$ inside various compact objects. Here solid, dotdashed and dashed lines are drawn for 4U 1820-30, PSR J1903+0327 and PSR J1614-2230 respectively. Horizontal red line corresponds to the value $\frac{4}{3}$.}
\label{Fig20}
\end{center}
\end{figure}

The adiabatic index ($\Gamma$) is an important thermodynamical quantity describing any instability inside a compact object which is defined as
\begin{equation}
\Gamma = \frac{\rho+p_r}{p_r}\frac{dp_r}{d\rho}. \label{Eq49}
\end{equation}
For stability the adiabatic index has to be greater than $\frac{4}{3}$ \cite{Heintzmann}. For our model we find that the value of $\Gamma$ is always larger than $\frac{4}{3}$ as evident from Fig.~\ref{Fig20}.

\section{Conclusions}\label{conc}
In this paper, we have obtained a class of solution of relativistic compact objects in spheroidal geometry having matter distribution anisotropic in nature obeying colour-flavor-locked (CFL) equation of state. The metric ansatz in spheroidal geometry was introduced by Vaidya \& Tikekar \cite{Vaidya} in which the $t=$ constant hypersurface has the geometry of a three-spheroid embedded in a four dimensional Euclidean space and stipulates a law of variation of density determined by the curvature of the physical 3-space. Assuming a density dependence on $B$-parameter according to Aguirre \cite{Aguirre}, the energy per baryon in CFL phase is evaluated which seems to take minimum value at zero external pressure as evident from Fig.~\ref{Fig2}. The minima of the plots increase with increasing mass of strange quark ($m_s$) and touches the line corresponding to the energy per baryon of $^{56}Fe$ ($\sim$ 930.4 $MeV$) for $m_s = 228.3$ $MeV$. Thus 3-flavour quark matter (SQM) in CFL phase is absolutely stable relative to $^{56}Fe$ for $m_s<228.3$ $MeV$. The stability of SQM increases with a decrease in the value of $m_s$. As for example when $m_s=150$ $MeV$, the energy per baryon $\sim$ 889.2 $MeV$ which represents weakly bound SQM compared to the value 856.1 $MeV$ when $m_s=50$ $MeV$. To apply the thermodynamically obatained EoS to some observed compact objects we have taken the compact objects 4U 1820-30, PSR J1614-2230, PSR J1903+0327, PSR J0030+0451 and PSR J0740+6620. It has been found that a wider range of $\lambda$ is possible for which CFL EoS may be obtained inside these compact objects. Apart from that PSR J1614-2230 and PSR J0740+6620 are found to be anis- otropic in nature with anisotropy parameter $\alpha=$ 0.39 and 0.45 respectively while the others show isotropic pressure distribution. The radius prediction of these compact objects suggests that CFL EoS gives much smaller radii compared to MIT bag EoS which can be noted from Table~\ref{tab1}. In CFL phase quarks form cooper pairs which collectively behave as boson and hence by virtue it does not need to obey Pauli's exclusion principle. Thus gravity can shrink the star to a more compact structure and as a result its radius decreases. From the mass-radius plots in Fig.~\ref{Fig7}, it is noted that CFL EoS puts a limit on the maximum mass which is $\sim$ 3.61 $M_\odot$ when $m_s = 0$ $MeV$ and decreases further for higher values of $m_s$. A comparison is made with MIT bag EoS which shows that CFL EoS allows higher value of maximum mass than MIT bag EoS for which $M_{max}=$ 2.03 $M_{\odot}$. Such high value of maximum mass permits our model to include wider range of compact objects which is also indicated in Fig.~\ref{Fig7}. From Fig.~\ref{Fig8}, it is evident that maximum central density permitted by CFL EoS is 1.221 $\times$ $10^{15}$ $gm/cm^3$. Above the maximum mass point it is noted that $\frac{\partial{M}}{\partial{\rho_0}}<0$, which is not allowed since this leads to instability of a fluid sphere \cite{Zeldovich}. From Figs.~\ref{Fig9}- \ref{Fig14}, it is obvious that all the necessary energy conditions are obeyed in this model. The equilibrium of the stellar configuration under mutual influence of graviational ($F_g$), hydrostatic ($F_h$) and anisotropic ($F_a$) forces are discussed interms of TOV equation given by Eq.~(\ref{Eq45}). Which shows that the combined effect of $F_h$ and $F_a$ balances $F_g$ so that total force is essentially zero at all points inside a star. The causality conditions and the inequality given by Abreu {\it et al.} \cite{Abreu} is satisfied as can be seen from Figs.~\ref{Fig16} - \ref{Fig19}. For stability against small adiabatic perturbations we calculate the value of adiabatic index $\Gamma$ from Eq.~(\ref{Eq49}) and study its radial variation which is plotted in Fig.~\ref{Fig20}. It is evident that at all interior points the value of $\Gamma$ is greater than $\frac{4}{3}$ showing that our model is stable against small perturbation for adiabatic flow.  
\begin{acknowledgements}
KBG is thankful for the fellowship provided by CSIR vide no. 09/1219(0004)/2019-EMR-I.
\end{acknowledgements}

\section{Declarations}
\begin{description}
\item[Funding:] A fellowship has been provided to K.B. Goswami by Council of Scientific and Industrial Research, India (vide no. 09/1219(0004)/2019-EMR-I)
\item[Conflicts of interest:] Not applicable 
\item[Availability of data and material:] This manuscript has no associated data
or the data will not be deposited, we have used only observed mass and radius of some known compact objects to construct
relativistic stellar models.
\end{description}

\end{document}